\documentclass[english,twocolumn]{revtex4-2}
\usepackage{babel}
\usepackage{amssymb}
\usepackage{graphicx}
\usepackage{color}
\usepackage{bm}
\usepackage{longtable}
\usepackage{amsmath} 
\usepackage{amsfonts}
\usepackage{amssymb}
\usepackage{mathtools}
\usepackage[T1]{fontenc}
\usepackage[latin1]{inputenc}
\usepackage{braket}
\usepackage{babel}%
\usepackage{graphicx}
\usepackage{color}
\usepackage{bm}
\usepackage{longtable}
\usepackage{amsmath}
\usepackage{amsfonts}
\usepackage{dsfont}
\usepackage{amssymb}
\usepackage{hyperref}
\newcommand{\ba}{\begin{eqnarray}}
\newcommand{\ea}{\end{eqnarray}}

\begin{document}
\title{Oblivious communication game, self-testing of projective and non-projective measurements and certification of randomness }
\author{A. K. Pan \footnote{akp@nitp.ac.in}}
\affiliation{National Institute of Technology Patna, Ashok Rajpath, Patna, Bihar 800005, India}
\begin{abstract}
We provide an interesting two-party parity oblivious communication game whose success probability is solely determined by the Bell expression. The parity-oblivious condition in an operational quantum theory implies the preparation non-contextuality in an ontological model of it. We find that the aforementioned  Bell expression has two upper bounds in an ontological model; the usual local bound and a non-trivial preparation non-contextual bound arising from the non-trivial parity-oblivious condition, which is smaller that the local bound. We first demonstrate the communication game when both Alice and Bob perform three measurements of dichotomic observables in their respective sites. The optimal quantum value of the Bell expression in this scenario enables us to device-independently self-test  the maximally entangled state and trine-set of observables, three-outcome qubit positive-operator-valued-measures (POVMs) and 1.58 bit of local randomness.  Further, we generalize the above communication game in that both Alice and Bob perform the same but arbitrary (odd) number ($n> 3$) of measurements. Based on the  optimal quantum value of the relevant Bell expression for any arbitrary $n$, we have also demonstrated device-independent self-testing of state and measurements.  
\end{abstract}
\pacs{03.65.Ta} 
\maketitle
\section{Introduction}

Bell theorem \cite{bell} is at the heart of quantum foundations. This no-go proof asserts that all quantum statistics of quantum theory cannot be accounted for by any ontological model satisfying locality. Later, it is found that Bell's theorem certifies the non-local correlation in a device-independent way in that  no characterization of devices is needed to assumed. Besides the immense impact of Bell's theorem on conceptual foundations of quantum theory, the device-independent quantum certification based on it leads a flurry of potential practical applications (see, for a review \cite{brunnerreview}) in quantum information processing task. 

Another pertinent no-go proof in quantum foundations - the Kochen and Specker (KS)\cite{kochen} theorem - proves an inconsistency between the quantum theory and non-contextual ontological models. While the demonstration of Bell's theorem requires two or more space-like separated systems, the KS theorem can also be demonstrated for a single system having dimension of the Hilbert space
$d \geq 3$. However, the traditional KS notion of non-contextuality is merely applicable to the deterministic ontological models of quantum theory and the ontic states strictly provide the values corresponding to the sharp projective measurements only.  The notion of KS non-contextuality was further generalized by Spekkens \cite{spek05} for positive-operator-valued-measures (POVMs) in any arbitrary operational theory. He also extended the formulation to the transformation and preparation non-contextuality. In the present work, the notion of preparation contextuality plays a important role.

The communication games \cite{ raz99,buhrman01,wolf02,bruk02,bruk04,bra03,amb08,buhrman10,opp,buhrman16,mar18, tava19, spek09,banik15,tava15,chill16,hameedi,him17, ghorai,ambanis19,saha19a,saha19b,tava2020} are widely used tools for studying the fundamental limiting features of an operational theory in terms of their ability to process information. In such a game, two or more parties jointly perform a given task with highest possible efficiency despite the amount and type of communication are constrained by some rules. In terms of the nature of communication from sender to receiver, there are two major classes of games; the one in which the dimension of communicated system in classical or quantum theory is bounded, and the other one, in which the obliviousness condition on the communication is imposed without any restriction on the dimension of the system and/or on the amount of communications. There is yet another class of games using energy constraints \cite{him17} and information content constraints \cite{tava2020}.  Both the classes of communication games can be played either in prepare-measure scenario or in entanglement assisted scenario \cite{spek09,banik15,tava15,chill16,hameedi,ghorai,ambanis19,saha19a,saha19b}.  The well-known parity-oblivious random access code  \cite{spek09,banik15,tava15,chill16,ghorai} is one of such communication games.

In this work, we provide an interesting two-party oblivious communication game  in which the sender (Alice) is allowed to communicate any amount of information but that should not reveal the parity information of the inputs to the receiver (Bob). We demonstrate the success probability of this parity-oblivious communication game is solely dependent on a suitable Bell expression. Note here that, obliviousness in an operational theory can equivalently be represented as the obliviousness at the level of ontic states if the ontological model of that operational theory is preparation non-contextual \cite{spek09}. In this connection, it is also worthwhile to note that, in two-input-two-output Bell scenario the preparation non-contextuality assumption in an ontological model of quantum theory can also be viewed as locality condition \cite{pusey18,uola20}. We may call it as trivial preparation non-contextuality condition. However, in a two-party Bell scenario beyond two-input-two-output  there could be certain forms of oblivious condition which can lead non-trivial restriction on the choices of inputs. In such a case the upper bound of the Bell expression may be reduced from the trivial case (the local bound), which we term here as non-trivial preparation non-contextual bound. Thus, for specific choice of states and measurements, it is possible that the optimal quantum value of the Bell expression may not be enough large to exhibit non-locality but the non-classicality in the form of non-trivial preparation contextuality may still be revealed. In quantum theory,  the optimal value of the Bell expressions enables one to self-test the state and measurements. 

Specifically, we propose an entanglement-assisted parity-oblivious communication game in that both Alice and Bob receive inputs $x,y\in \{1,2,.....n\}$ with odd $n\geq 3$ and according to which they perform local measurements on their respective sites. Each of the local measurements produces dichotomic outputs $a,b\in \{0,1\}$. The inputs of Alice satisfy a parity-oblivious condition and this in turn provides  that a functional relationship between Alice's observables has to be satisfied. We show that the success probability of the communication game is solely determined by the value of a family of Bell expressions (say, $\mathcal{B}_{n}$) which has a local bound and a non-trivial preparation non-contextual bound. We demonstrate that optimal quantum value $(\mathcal{B}_{n})_{Q}^{opt}$  enables to device-independently self-test the entangled state and a set of projective measurements. 

We first demonstrate the communication game for $n=3$ which allows us to self-test a trine-set of observables and entangled state. We then show that a simple modification of the aforementioned game can certify the three-outcome qubit POVMs which in turn can be used certify $1.58$ bit of local randomness. Further, we generalize the aforementioned three-input game to any (odd) arbitrary $n$ input game and optimal quantum success probability enables the self-testing of maximally entangled state and a set of observables. We further discuss that such a generalization does not enable us to certify the randomness. 

The plan of the paper is the following. In Sec. II, we provide the preliminaries of oblivious communication game, the notion of preparation non-contextuality in an ontological model, the self-testing protocols and the device-independent randomness certification.  We provide a specific entanglement-assisted parity-oblivious game in which Alice and Bob perform three measurements each and optimization of the success probability of that game in Sec. III. In Sec. IV, we provide the self-testing protocol that certifies the entangled state and the trine-set of observables. The self-testing of three-outcome POVMs and local randomness is provided in Sec. V. The generalization of the communication game for any arbitrary odd $n$ is provided in Sec. V. We summarize our results in Sec. VI. 

\section{preliminaries}
Before presenting the main results, we briefly summarize the notion of preparation non-contextuality in an ontological model, the parity oblivious communication game, the device-independent self-testing and certifications randomness.  
\subsection{Operational theory and ontological model}
 We invoke an elegant framework of an ontological model \cite{hari,spek05} of quantum theory to introduce the notion of non-contextuality from modern perspective.  Given a preparation procedure $P$ and a measurement procedures $M$, an operational theory assigns probability $p(k|P, M)$ of obtaining a particular outcome $k$.   In quantum theory, a preparation procedure $(P)$ produces a density matrix $\rho$ and measurement procedure $(M)$ (in general described by POVMs $(E_k$)) provides the probability of a particular outcome $ k $ is given by $p(k|P, M)=Tr[\rho E_{k}]$- the Born rule. 

In an ontological model of quantum theory, it is assumed that whenever $\rho$ is prepared by $P$, a probability distribution $\mu_{P}(\lambda|\rho)$ in the ontic space $\Lambda$ is prepared, satisfying $\int _\Lambda \mu_{P}(\lambda|\rho)d\lambda=1$ where $\lambda \in \Lambda$. The probability of obtaining an outcome $k$ is given by a response function $\xi_{M}(k|\lambda, E_{k}) $ satisfying $\sum_{k}\xi_{M}(k|\lambda, E_{k})=1$ where a measurement operator $E_{k}$ is realized through $M$. A viable ontological model should reproduce the Born rule, i.e., $\forall \rho $, $\forall E_{k}$ and $\forall k$, $\int _\Lambda \mu_{P}(\lambda|\rho) \xi_{M}(k|\lambda, E_{k}) d\lambda =Tr[\rho E_{k}]$.

An ontological model of an operational theory can be assumed to be non-contextual in the following way \cite{spek05};  if two experimental procedures are equivalent in operational theory then they can be represented non-contextually in an ontological model.   Then, an ontological model of quantum theory is assumed to be preparation non-contextual if $\forall M, \  k:$
\begin{align}
\label{ass}
  p(k|P, M)=p(k|P^{\prime},M)	\Rightarrow \mu_{P}(\lambda|\rho)=\mu_{P^{\prime}}(\lambda|\rho)
	\end{align}
 where the $\rho$ is prepared by two distinct preparation procedures $ P $ and $ P^{\prime}$ \cite{spek05, pan19,panep21}. We shall shortly see that in a preparation non-contextual ontological model the parity-oblivious constraint in a communication game in operational quantum theory implies equivalent  obliviousness condition at the level of  ontic states. 

\subsection{Oblivious communication games}
Consider a scenario where two distant parties, Alice and Bob, collaborate to perform a common task through a one-way communication \cite{
raz99,buhrman01,wolf02,bruk02,bruk04,bra03,buhrman10,opp,buhrman16,mar18, tava19}.  Alice (Bob) receives an input $x\in \{1, . . . , n_{A}\}$ ($y\in \{1, . . . , n_{B}\}$) with probability distribution $p_{A}(x)$ ($p_{B}(y)$).  Bob's task is to guess a function of their interest $f(x, y)$ with the help of Alice's communication. For this, he encodes his answer in an output variable, say, $b \in \{0,1\}$. Let $p(b|x, y)$ represents the probability of obtaining an binary output $b$ given inputs $x, y$. The input may also contain the output of Alice. The guessing probability of the function $f(x, y)$, as a linear function of the observed probabilities $\{p(b|x, y)\}$. Thus, any linear figure of merit can
be expressed as, 
\begin{align}
	\mathbb{P} = \sum\limits_{x,y} \mathcal{C}_{x,y}^{b} p_{A}(x) p_{B}(y) p(b = f(x, y)|x, y)
\end{align}

where $\mathcal{C}_{x,y}^{b}$ is the pay-off function of the game quantifies  the
normalized weightage for guessing the correct $f(x, y)$.  The quantum advantage of a communication game over classical resources becomes trivial if Alice is allowed to send her input $x$ to Bob. However, if some constraints are imposed on the communication from Alice to Bob, then the supremacy of quantum resources may be exhibited. One of such constraints can be bounding the dimension of the input. Another one, in which we are particularly interested here is the parity-obliviousness condition \cite{spek09,banik15,tava15,chill16,hameedi,ghorai,ambanis19,saha19a,asmita19,saha19b,mahato}. Such a condition implies that there is no restriction on the number of communication but that should not covey the information about a particular property of the inputs. 

Let in an operational theory, Alice prepares the inputs $x$ by the preparation procedures $P_{x}$ and upon receiving the input $y$, Bob performs the measurement of $M_{y}$. Consider that there are $L$ subsets having same number of elements of the input $P_{l}\subset P_{x}$ with $l=1,2,3 . . ..L$. An oblivious condition demands that given an input is not distinguishable whether it has come from $P_{l}\subset P_{x}$ or from $P_{l^{\prime}}\subset P_{x}$ even when  Alice's communication is not restricted. For our purpose it will be enough to consider the input of Alice is uniformly distributed so that $p_{A}(x)=1/|P_{x}|$ where $|P_{x}|$ is the cardinality of the set. Then, for an oblivious game $	\forall \ l, l^{\prime}, y, b $ we can write

\begin{align}
\label{po11}
 \sum\limits_{P_{x}\in P_{l}} p(P_{x}|b,M_{y}) = \sum\limits_{P_{x}\in P_{l^{\prime}}} p(P_{x}|b,M_{y}) 
\end{align}
Using the Bayes rule one can write $p(P_{x}|b,M_{y})=p(b|P_{x},M_{y}) p(x,y)/p(b|M_{y})$. By noting $p(x,y)=p_{A}(x)p_{B}(y)$  the Eq. (\ref{po11}) can be written as  
 
\begin{align}
\label{po111}
\forall b, y \ \ \sum\limits_{P_{x}\in P_{l}} p(b|P_{x},M_{y}) = \sum\limits_{P_{x}\in P_{l^{\prime}}} p(b|P_{x},M_{y}) 
\end{align}
for $	\forall \ l, l^{\prime}, y, b $. This means that the two input sets $P_{l}$ and $P_{l}^{\prime}$ cannot be distinguished by any outcome $b$ and any measurement $M_{y}$ in an operational theory. This takes the form  of the premise of the notion of preparation non-contextuality given in Eq. (\ref{ass}). Assuming preparation non-contextuality in an ontological model of the above operational theory we can write  
\begin{align}
\label{pnc}
 \sum\limits_{{P_{x}}\in P_{l}} \mu(\lambda|P_{x}) =  \sum\limits_{{P_{x}}\in P_{l^{\prime}}} \mu(\lambda|P_{x}) 
\end{align}
where $\lambda \in \Lambda$ is the ontic state and $\Lambda$ is the ontic state space. Using Bayes rule once again it can be shown that 
\begin{align}
\label{pnc}
 \sum\limits_{{P_{x}}\in P_{l}} \mu(P_{x}|\lambda) =  \sum\limits_{{P_{x}}\in P_{l^{\prime}}} \mu(P_{x}|\lambda) 
\end{align}
which implies that for preparation noncontextual models, the satisfaction of obliviousness condition in an operational theory provide equivalent representation at the level of the ontic states. In other words, the obliviousness condition must be satisfied at the level of ontic states $\lambda$ too for the preparation non-contextual model. In this work, we consider a particular obliviousness condition - the parity-obliviousness one - in which no parity information of the inputs will be transmitted to Bob due to Alice's communication.  Similarly, for preparation noncontextual ontological models, the ontic state $\lambda$ cannot contain any information about the parity.

\subsection{Device-independent Self-testing }
Self-testing in its traditional form is a device-independent protocol that aims to uniquely characterize the nature of the target quantum state and measurements solely from the correlations. Essentially, this requires finding a suitable Bell inequality whose maximum violation is achieved uniquely by the target state and measurements involved.  Given the communication game discussed above, the observed joint probability in quantum theory can be obtained from Born rule is given by  $p(a b|x,y)=Tr[\rho_{AB}(A_{a|x}\otimes B_{b|y}]$ where $\rho_{AB}=|\psi_{AB}\rangle\langle\psi_{AB}|$ is an entangled state and $\{A_{a|x}\}$and $ \{B_{b|y}\}$ are the set of local measurements belong to Alice and Bob respectively. 

The aim of the self-testing is to find the suitable correlations which can uniquely be realized by the target state and measurements. The traditional self-testing scenario first proposed by Mayers and Yao \cite{mayer98}. Later, McKague and Mosca \cite{mckague} used this isometric embedding to develop a generalized Mayers-Yao test \cite{mayers}. Since then, a flurry of works on this topic has been reported \cite{ mckague12,wu,mckague16,and17,cola,supic18,bowels18prl,bowles18,coopmans,tavakoli19a}. Related works, such as, certification of binary outcomes has also been reported \cite{quin19}. Another interesting proposal for device-independent self-testing of Pauli observables is put forwarded in \cite{bowels18prl, bowles18} by using three CHSH inequalities.  For a recent review, we refer Ref. \cite{supicrev}. 

However, although device-independent scenario uses minimal assumptions, conclusive experimental certification are challenging. To circumvent this issue, semi-device-independent self-testing protocols  has been proposed \cite{tavakoli18,smania,mir19,gomez16,gomez18,tavakoli20,mir19,farkas,kartik,paw20} where the entanglement is not required and the dimension of the system are known. Such protocols are claimed to be more appealing for experimentalists compared to fully device-independent Bell test.  In this work, based on the optimal quantum success probability of a suitable communication game, we provide schemes to device-independently self-test the entangled state, a specific set of projective measurements and the three-outcome extremal qubit POVMs. 

\subsection{Certification of randomness}
Randomness is a powerful resource having wide field of applicability ranging from scientific research to our daily life. Classical algorithms, whatever powerful it may be, can only produce psudo-random number, whose unpredictability relies on the complexity of the generator \cite{matsumoto}.  On the other hand, quantum theory provides intrinsic randomness through the unpredictability of the Born rule.   Device-independent randomness generation rely on a fundamental relation between the non-locality of quantum theory and its random character which is usually expressed in terms of a trade-off between the probability of guessing correctly the outcomes of measurements performed on quantum systems and the amount of violation of a given Bell inequality \cite{bell, acin07, brunnerreview}. 

Such a strategy of certifying device-independent randomness was first put forwarded by Colbeck \cite{colbeck06} in his PhD thesis.  Adopting the similar strategy in \cite{pironio} the relation between randomness and violation of Bell's inequality is established through the non-local guessing games. The joint probability $P(a b|x,y)$ can be obtained when Alice and Bob perform measurements according to the given inputs. In our case, we have  inputs $x,y\in{1,2...n}$ and outputs $a,b\in\{0,1\}$. Then there will be $2n^{2}$ number of joint probabilities, can be viewed as a component of a vector ${\bf P}= \{p(a,b|x,y)\}$ is referred to as behavior which characterizes the systems of Alice and Bob \cite{nieto}.  Here is our communication game, it is assumed that $\bf{P}$ is given which means it is a promise on the behavior.  In the non-local guessing game there is another party, Eve, whose goal is to guess the Alice's outcome for a certain input (say, $x^{\ast}$) with highest possible probability. A strategy of Eve can be that she prepares the quantum state $|\Psi_{ABE}\rangle\in \mathcal{H}_{A}\otimes \mathcal{H}_{B}\otimes \mathcal{H}_{E}$ for Alice and Bob, so that  $|\Psi_{AB}\rangle$ can be obtained by tracing out her system. Given inputs $x$ and $y$ Alice and Bob measures a set of POVMs $\{A_{a|x}\}$ and $\{B_{b|y}\}$. Thus, $P(a b|x,y)=Tr[(A_{a|x}\otimes B_{b|y}\otimes \mathbb{I})\rho_{ABE}]$. Given a special input $x^{\ast}$, the local guessing probability can be written as

\begin{align}
	G=\max_{F} P(a,a|A_{x^{\ast}}, F)=\max_{F} Tr[(A_{a|x^{\ast}}\otimes \mathbb{I}\otimes F_{a})\rho_{ABE}]
\end{align}
where $F=\{F_a\}$ is the POVM of Eve whose measurement result provides her best guess of Alice's outcome. The min-entropy can be used a measure of randomness so that $H_{min}(a|x)=-log_{2} 	G$  amount of randomness is generated by Alice. 

Since then quite number number of works along this direction have been reported \cite{colbeck, acin12,pir13,bancal,acin16,gomez18, and18,curchod} and verified experimentally \cite{Yang,Peter,zhang20}. Note that, the device-independent randomness certification faces practical challenges appears with the loophole-free violation of Bell inequality by lowering the bit rate.  In recent times, loophole-free Bell tests have been realized \cite{lhf1,lhf2,lhf3} which in turn enables experimental demonstrations of device independent random number certification \cite{Yang, Peter}.  However, such implementation still remains a difficult task to perform commonly. To tackle this practical issue, the device-independent self-testing of random number generator in a prepare-measure scenario is proposed in \cite{lungi}. Semi-device-independent randomness certification protocols \cite{li11,li12,Wen,pan21} in a prepare-and-measure scenario have also been proposed where the dimension of the quantum system is known. Here we demonstrate the device-independent certification of randomness using the optimal success probability of our communication game.

\section{A parity-oblivious communication game and relevant Bell's inequality}
Equipped with the preliminary ideas about ontological model and oblivious communication game, we are now in a position to introduce a specific parity-oblivious communication game. We first provide a parity-oblivious communication game where Alice and Bob hold input $x,y \in \{1,2,3\}$ and outputs are $a,b\in \{0,1\}$. This corresponds to the measurements dichotomic observables of $A_{x}$ and $B_{y}$ by Alice and Bob respectively.  Using her output, Alice prepares six input states $x^{i}\in (x,a)\equiv \{10,11,20,21,30,31\}$ where $i=1,2...6$. For our purpose we consider uniform distribution of inputs of Alice and also for Bob, so that,  $p_{A}(x)=p_{B}(y)=1/3$. The winning rule of the game is that Bob's output must be $b=\delta_{x,y}\oplus_{2}a$. In an operational theory the success probability of this communication game is 
\begin{align}
	\mathbb{P}_{3} = \dfrac{1}{9}\sum\limits_{x,y=1}^{3} p(b=\delta_{x,y}\oplus_{2}a)|x, y)
\end{align}

When there is no restriction on the inputs, the success probability can be cast as
\begin{align}
	\mathbb{P}_{3}=\frac{1}{2}\left(1+\frac{\langle \mathcal{B}_{3}\rangle}{9}\right)
\end{align}

where 
\begin{eqnarray}
\label{bi}
	\mathcal{B}_{3}&=& A_{1}\otimes(-B_{1} +B_{2} +B_{3}) \\
	\nonumber
	&+&  A_{2}\otimes(B_{1} -B_{2} +B_{3})+A_{3}\otimes(B_{1} +B_{2} -B_{3})
\end{eqnarray}

The correlation $\langle A_x B_x\rangle=\sum_{a,b} (-1)^{a\oplus_{2}b}P(a b|x,y)$. The local bound of the Bell expression is $	(\mathcal{B})_{local}\leq 5$. 

We now impose the parity-oblivious restriction on communication. Consider the input set is divided into two subsets having equal number of elements; the even parity set $P_{l}=\{x^{i}: x\oplus_{2} a=0\}$ and the odd parity set $P_{l^{\prime}}=\{x^{i}: x\oplus_{2} a=1\}$. The parity-obliviousness condition demands that
\begin{align}
	\sum\limits_{x^{i}\in P_{l}} p(b|x^{i},y)=\sum\limits_{x^{i}\in P_{l^{\prime}}} p(b|x^{i},y)
\end{align}
As already mentioned, the parity obliviousness in an operational theory implies similar consequence at the level of ontic states if the ontological model is preparation non-contextual.  

In quantum theory, Alice encodes her input string of $x^{i}$ into pure quantum states $\rho_{x^{i}}$ prepared by a procedure $P_{x^{i}}$. Bob performs a two-outcome measurement $B_{y}$ for for every $y\in\{1,2,3\}$ and reports outcome $b$ as his output. If Alice and Bob share an entangled state $|\psi_{AB}\rangle$ then Alice can steer the states $x^{i}$ to Bob by measuring three dichotomic observables $A_{x}$ on her particle corresponding to the input $x\in \{1,2,3\}$. For example, $ \rho_{11}=Tr_{A}[\left(\Pi_{A_1}^{+}\otimes \mathbb{I}\right)\rho_{AB}\left(\Pi_{A_{1}}^{+}\otimes\mathbb{I}\right)]/Tr[\rho_{AB}\left(\Pi_{A_{1}}^{+}\otimes\mathbb{I}\right)]$ where $\Pi_{A_{1}}^{+}=(\mathbb{I}+A_{1})/2$ is the projector of the Alice's observable $A_1$. Also, $ \rho_{21}=Tr_{A}[\left(\Pi_{A_2}^{-}\otimes \mathbb{I}\right)\rho_{AB}\left(\Pi_{A_{2}}^{-}\otimes\mathbb{I}\right)]/Tr[\rho_{AB}\left(\Pi_{A_{2}}^{-}\otimes\mathbb{I}\right)]$ and $ \rho_{31}=Tr_{A}[\left(\Pi_{A_3}^{+}\otimes \mathbb{I}\right)\rho_{AB}\left(\Pi_{A_{3}}^{+}\otimes\mathbb{I}\right)]/Tr[\rho_{AB}\left(\Pi_{A_{3}}^{+}\otimes\mathbb{I}\right)]$. Note that, $\rho_{x1}+\rho_{x0}=\mathbb{I}$ with $x=1,2,3$.  The parity-oblivious condition in quantum theory reads as 
\begin{align}
\label{poqm}
	\sum\limits_{x^{i}|x\oplus_{2} a=0} \rho_{x^{i}}=	\sum\limits_{x^{i}|x\oplus_{2} a=1} \rho_{x^{i}}
\end{align}
This explicitly means $\rho_{11}+\rho_{20}+\rho_{31}=\rho_{10}+\rho_{21}+\rho_{30}$. It is straightforward to see that the parity-oblivious condition given by Eq. (\ref{poqm}) provides a non-trivial functional relation  $\sum\limits_{x=1}^{3} A_{x}=0$ between the observables that has to be satisfied  in quantum theory. 

Equivalently, in an ontological model of quantum theory, the preparation non-contextuality assumption provides

\begin{align}
	\sum\limits_{x^{i}|x\oplus_{2} a=0}\mu(\lambda|P_{x^{i}})=	\sum\limits_{x^{i}|x\oplus_{2} a=1}\mu(\lambda|P_{x^{i}})
\end{align}
In an preparation non-contextual ontological model the equivalent condition of $\sum\limits_{x=1}^{3} A_{x}=0$ needs to be used to derive the upper bound of the Bell expression $\mathcal{B}_{3}$. Imposing this non-trivial condition in an ontological model, the local bound of $\mathcal{B}_{3}$ gets reduced to the non-trivial preparation non-contextual bound $	(\mathcal{B}_{3})_{pnc}\leq 4$. Importantly, the choices of Alice's observable optimize the quantum value of $(\mathcal{B}_{3})_{Q}=6$ satisfies the parity oblivious condition in quantum theory.

In order to derive optimal quantum value of the Bell expression $\mathcal{B}_{3}$,  we use sum-of-square (SOS) approach \cite{bamps}, so that $(\mathcal{B}_{3})_{Q}\leq \beta_{3}$  for all possible quantum states and measurement operators $A_{x}$ and $B_{y}$  where  $\beta_{3}$ is the upper bound on the quantum  value of $(\mathcal{B}_{3})_{Q}$. This is equivalent to showing that there is a positive semi-definite operator $\gamma_{3} \geq 0$, that can be expressed as $\langle \gamma_{3}\rangle_{Q}=\beta_{3} -(\mathcal{B}_{3})_{Q} $. This can be proven by considering a set of suitable positive operators $L_{y}$ which is polynomial functions of   $A_{x}$ and $B_{y}$, so that 

  \begin{align}
\label {gamma}
	 \gamma_{3} =\frac{1}{2} \sum_{y=1}^{3} \omega_{y} L_y^\dagger L_y 
 \end{align}
	where $ L_y $'s are positive operators. For the Bell expression given by Eq. (\ref{bi}), we choose the operators $L_{y}$ as

\begin{align}
\label{mi}
	L_{y}|\psi\rangle=\frac{1}{\omega_{y}}\left(\sum\limits_{x=1}^{3} \alpha^{x,y}_{3} A_{x}\right) |\psi\rangle -B_{y} |\psi\rangle
\end{align}
where $\alpha^{x,y}_{3}=1 (-1)$ when $x\neq y  (x=y)$. Also, 
\begin{align}
\label{omega}
			\omega_{y} =||\sum\limits_{x=1}^{3} \alpha^{x,y}_{3} A_{x}|\psi\rangle||
\end{align}
 where $||.||$ is the Euclidean norm of a vector.  Plugging Eq. (\ref{mi}) into Eq. (\ref{gamma}) and by noting that $A_{x}^{\dagger} A_{x}=B_{y}^{\dagger} B_{y}=\mathbb{I} $, we get

\begin{align}
\langle \gamma_{3}\rangle_{Q}=-(\mathcal{B}_{3})_{Q} + \sum\limits_{y=1}^{3}\omega_{y}
\end{align}
which can be re-written as 
\begin{align}
(\mathcal{B}_{3})_{Q} =\sum\limits_{y=1}^{3}\omega_{y} -\langle \gamma_{3}\rangle_{Q} 
\end{align}
In order to maximize $(\mathcal{B}_{3})_{Q}$, we write
\begin{align}
\label{max}
max\left((\mathcal{B}_{3})_{Q}\right) \leq max\left(\sum\limits_{y=1}^{3}\omega_{y}\right) +max\left(-\langle \gamma_{3}\rangle_{Q}\right) 
\end{align}

We separately derive $max\left(\sum\limits_{y=1}^{3}\omega_{y}\right) $ and $max\left(-\langle \gamma_{3}\rangle_{Q}\right)$ \ \ $\forall \psi, A_{x}, B_{y}$. To maximize $\sum\limits_{y=1}^{3}\omega_{y}$, we use the concavity inequality \cite{brad} $	\sum\limits_{y=1}^{3}\omega_{y}\leq \sqrt{3 \sum\limits_{y=1}^{3} (\omega_{y})^{2}}$. From the definition of $\omega_{y}$ in Eq. (\ref{omega}) we can write $(\omega_{1})^{2}= \langle\psi|(-A_{1}+A_{2} +A_{3})^{2} |\psi\rangle = 3 + \langle\psi|\left(-\{A_1,A_2\} +\{A_2,A_3\}-\{A_1,A_3\}\right)|\psi\rangle$. The quantities $(\omega_{2})^{2}$ and $(\omega_{3})^{2}$ can also be written in the similar manner. Using them, we have 
\begin{align}
\label{dd}
	\sum\limits_{y=1}^{3} \omega_{y}\leq \sqrt{3\left(9  - \langle \Delta_{3} \rangle \right)}
	\end{align}
where the quantity $\Delta_{3}$ is explicitly written as 
\begin{align}
\label{mm}
	\Delta_{3}= \{A_1,A_2\} +\{A_2,A_3\}+\{A_1,A_3\}
\end{align}
Here $\{ \}$ denotes anticommutation.  This means that minimizing $\langle  \Delta_{3}\rangle$ provides $	max\left(\sum\limits_{y=1}^{3} \omega_{y}\right)$.

For dichotomic observables satisfying $A_{x}^{2}=\mathbb{I}$, by considering $|\psi^{\prime\rangle}=\left(A_{1}+A_{2}+A_{3}\right)|\psi\rangle$ (where $|\psi\rangle$ is a non-zero vector), we can write 
\begin{align}
\label{dd1}
\langle\Delta_{3}\rangle =-3  +  \langle \psi^{\prime}|\psi^{\prime}\rangle
\end{align}

Note that the inner product $\langle \psi^{\prime}|\psi^{\prime}\rangle$ is in general non-negative. It becomes zero only when $|\psi^{\prime}\rangle$ is zero vector. The minimum value of $\langle\Delta_{3}\rangle$ is obtained when the inner product is zero. But since $|\psi\rangle$ is non-zero then $A_{1}+A_{2}+A_{3}=0$  needs to be satisfied. We obtain $	max\left(\sum\limits_{y=1}^{3} \omega_{y}\right) =6$. This also ensures that each of the $\omega_{y}$s is equal to 2 thereby implying $\langle \{A_x,A_x^{\prime}\}\rangle=-1$ for $x\neq x^{\prime}$. 

We now consider $max\left(-\langle \gamma_{3}\rangle_{Q}\right) $ $\forall \psi, A_{x}, B_{y}$.  Since $\gamma_{3}$ is a positive operator, $max\left(-\langle \gamma_{3}\rangle_{Q}\right)=0 $ for any $|\psi\rangle$ This means $\langle\psi L_{y}^{\dagger}L_{y}|\psi\rangle=0$, and consequently $L_{y}|\psi\rangle=0$, i.e.,

\begin{eqnarray}
\label{conda}
	\sum\limits_{x=1}^{3} \alpha^{x,y}_{3} A_{x}|\psi\rangle= \omega_{y} B_{y} |\psi\rangle
\end{eqnarray} 

Putting altogether, from Eq.(\ref{max}), we thus have the optimal value $(\mathcal{B}_{3})_{Q}^{opt} =6$.
It can be found from Eq. (\ref{conda}) that Bob's observables satisfy the relation $B_{y}=-A_{x}$ when $x=y$. This in turn provides the success probability $(\mathbb{P}_{3})^{opt}_{Q}=5/6$ compared to the non-trivial preparation non-contextual bound  $(\mathbb{P}_{3})_{pnc}=13/18$.  

One of the choices of Alice's of observables can even be found for qubit system are given by
\[A_{1}=\sigma_{z};  \ \  A_2=\frac{\sqrt{3}}{2}\sigma_{x}-\frac{1}{2}\sigma_{z};  \ \  A_3=-\frac{\sqrt{3}}{2}\sigma_{x}-\frac{1}{2}\sigma_{z}\]
which are the trine-spin axes satisfying $a_{x}.a_{x\prime}=-1/2$ for $x\neq x^{\prime}$ where $a_{x}$ is the Bloch vector. As already mentioned that from  Eq. (\ref{conda}) one finds $B_{y}=-A_{x}$ if $x=y$. This provides $\langle A_{x}\otimes B_{y}\rangle=-1 (1/2)$ when $x=y$ ($x\neq y$) and the state required for obtaining the optimal  value is 

\begin{align}
\label{enstate}
	|\phi^{+}\rangle=\frac{1}{\sqrt{2}}\left(|00\rangle+|11\rangle\right)
\end{align}

The optimal quantum value of the Bell expression $\mathcal{B}_{3}$ enables us to self-test the entangled state and projective measurements of trine-set of observables from the observed statistics. In the following, we provide the device-independent self-testing protocols based on  $\mathcal{B}_{3}^{opt}$. 
\section{self-testing of state and trine-set of observables }
 As mentioned earlier, for self-testing of state and measurements, one requires correlations $p(ab|x,y)$ which can be reproduced uniquely by the state and measurements (upto a certain equivalence class). Hence the target state and measurements can be certified from the correlation alone. In other words, the self-testing technique implies the existence of local unitaries along with axillary systems so that the target state and measurements can be inferred from the physical state and measurements. In our scenario when Alice and Bob performs three measurements each, we can find that the measurements can be expressed using only real numbers. However, we shall see when Alice and Bob performs more than three measurements each, the observables required for achieving optimum quantum value of Bell expressions cannot be expressed using the real numbers only. In such a case the correlations are invariant under complex conjugation or transposition \cite{mckague, bowles18,supicrev}. Since transpose is not a valid unitary map, the self-testing protocol needs to be suitably modified in this case \cite{mckague, bowles18,supicrev}, which is provided in Sec. VI. 

Let us first provide the self-testing protocol based on the optimal value $(\mathcal{B}_{3})_{Q}^{opt}$. The measurement can be considered projective as according to Naimark dilation theorem any non-projective measurements can be considered as projective measurement over a dilated Hilbert space. For our purpose, we invoke the SWAP circuit scheme \cite{mckague,mayer98,mayers} to demonstrate that the optimal quantum value $(\mathcal{B}_{3})_{Q}^{opt}$ implies the existence of an isometry $\Phi$ so that $\Phi: \mathcal{H}_{A}\otimes \mathcal{H}_{B} \rightarrow (\mathcal{H}_{A}\otimes \mathcal{H}_{A^{\prime}})\otimes(\mathcal{H}_{B}\otimes\mathcal{H}_{B^{\prime}})$ and $|\psi\rangle_{AB}\rightarrow|\chi\rangle_{AB}\otimes |\phi^{+}\rangle_{A^{\prime}B^{\prime}}$. Here, prime and non-prime denote the  reference and physical systems respectively. 

In order to find the self-testing properties, let us define the observables $Z_{A}= A_1$, ${X}_A = (A_3 -A_2)$, $Z_B = -B_1 $ and ${X}_B = (B_2 -B_3)$. Further we define  $\tilde{X}_A = X_{A}/||X_{A}||= (A_3 -A_2)/\sqrt{3}$ $\tilde{X}_B = X_{B}/||X_{B}||= (B_2 -B_3)/\sqrt{3}$. We derive the following properties.

\begin{align}
\label{st1}
	Z_A |\psi\rangle_{AB}={Z}_B|\psi\rangle_{AB};  \  \tilde{X}_A |\psi\rangle_{AB}=\tilde{X}_B|\psi\rangle_{AB}\\
\label{st2}
	\{Z_A, \tilde{X}_A\} |\psi\rangle_{AB} =\{{Z}_B, \tilde{X}_B\} |\psi\rangle_{AB}=0
	\end{align}
	
Details of the derivation of Eqs. (\ref{st1}) and (\ref{st2}) can be found in Appendix A. 

\begin{figure}[h]
  \includegraphics[width=8 cm,height=5 cm]{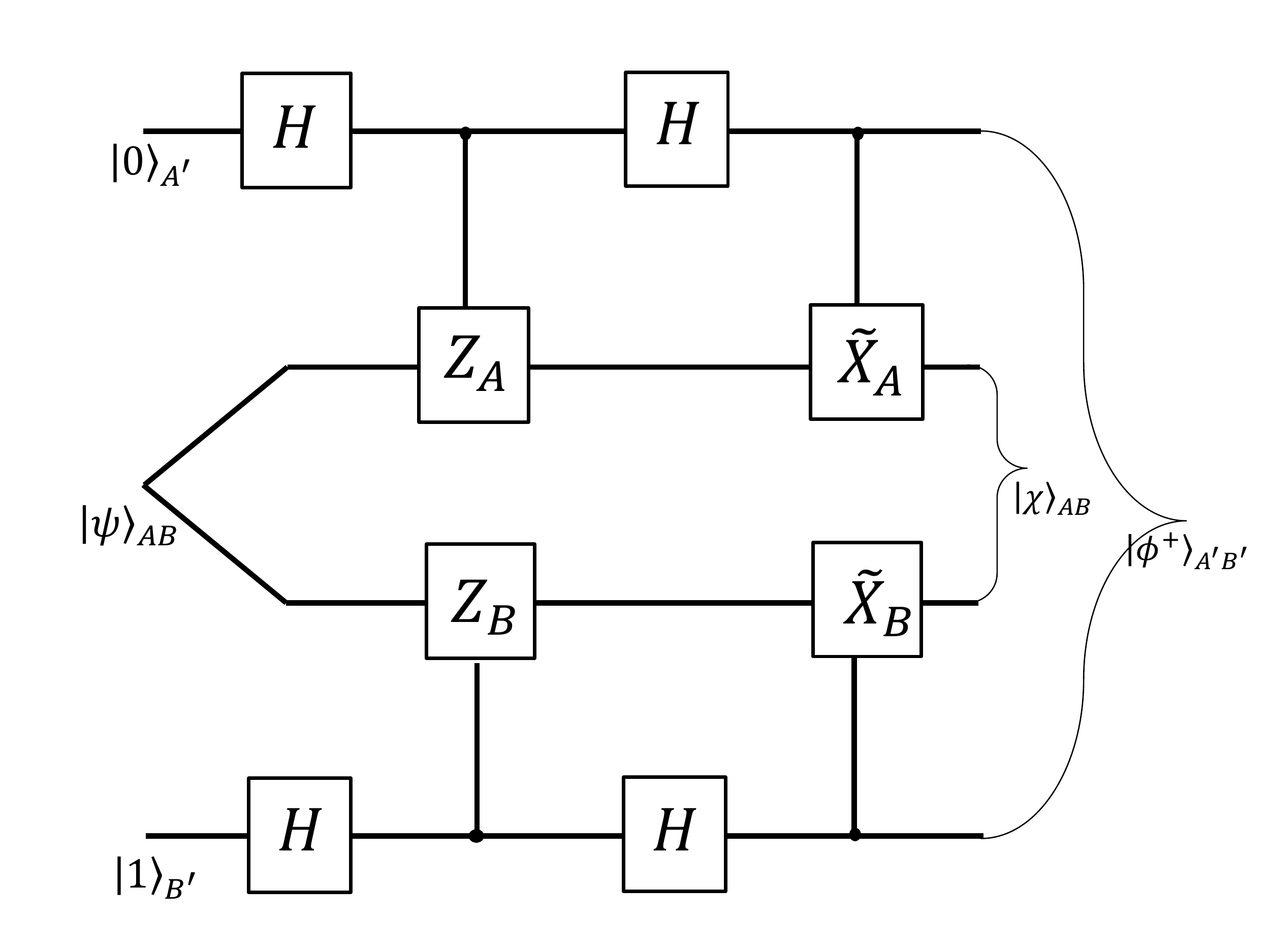}
 \centering
  \caption{(color online) Circuit based on the Bell's inequality in Eq. (\ref{bi}) for self-testing of two-qubit maximally entangled state and trine-set  of measurements. The operations ${Z}_A$, ${Z}_B$, $\tilde{X}_B$ and $\tilde{X}_B$ are defined in the main text and $H$ is the Hadamard gate. Detailed derivation is placed in Appendix A. 
 }
  \label{fig1}
\end{figure}
If $|\psi\rangle_{AB} \in \mathcal{H}_{A}\otimes\mathcal{H}_{B}$ is the state and $A_x\in \mathcal{H}_{A}$ and $B_y\in \mathcal{H}_{B}$ (where $x,y=1,2,3$) are the observables providing the optimal value of $\mathcal{B}_{3}$ then from the circuit  given by Fig.1, it can be proved that there exist local unitary operation $\Phi$ and ancila state $|00\rangle_{A^{\prime}B^{\prime}}$, such that, 

\begin{align}
\label{sf11}
\Phi(|\psi\rangle_{AB}\otimes |00\rangle_{A^{\prime}B^{\prime}})=|\chi\rangle_{AB} \otimes |\phi^{+}\rangle_{{A^{\prime}B^{\prime}}}
\end{align}
where $|\chi\rangle_{AB}=\frac{(1+ Z_A)}{\sqrt{2}}|\psi\rangle_{AB}$ is the so-called `junk' state. This is obtained by using the self-testing properties given by Eqs. (\ref{st1}) and (\ref{st2}). Corresponding to the measurements using the SWAP the circuit one gets
 
\begin{eqnarray}
\label{sf12}
	&&\Phi(B_{y}|\psi\rangle_{AB}\otimes |00\rangle_{A^{\prime}B^{\prime}})=|\chi\rangle \otimes (\mathbb{I}\otimes B_{y}^{\prime})|\phi^{+}\rangle_{{A^{\prime}B^{\prime}}}\\
	\label{sf13}
	&&\Phi(A_{x}|\psi\rangle_{AB}\otimes |00\rangle_{A^{\prime}B^{\prime}})=|\chi\rangle \otimes (A_{x}^{\prime}\otimes \mathbb{I})|\phi^{+}\rangle_{{A^{\prime}B^{\prime}}}\\
\label{sf14}
		&&\Phi(A_{x}B_{y}|\psi\rangle_{AB}\otimes |00\rangle_{A^{\prime}B^{\prime}})=|\chi\rangle \otimes (A_{x}^{\prime}\otimes B_{y}^{\prime})|\phi^{+}\rangle_{{A^{\prime}B^{\prime}}}
		\end{eqnarray}
		The details of the calculation is given in the Appendix A. This thus demonstrates that the self-testing protocol based on the Bell expression $\mathcal{B}_{3}$ given in Eq. (\ref{bi}) provides the equivalence between the reference and physical experiments. This in turn device-independently certify the maximally entangled state and trine-set of observables.

\section{Certifying three-outcome POVMs and randomness}

 In this section, we first argue how a simple modification of the earlier game and optimal quantum value in that case certify a three-outcome extremal qubit POVM and more than one bit of local randomness. For this we keep the original game as it is but introduce an additional input $x=4$ to Alice. This can be used to certify the three-outcome POVMs. Let us first explain the scenario. Alice receives an additional input $x=4$ and according to which she performs the measurement of a three-outcome extremal POVMs, say,  $A_{4}=\{A_{k|4}\}$ where $k=1,2,3$. This means each of the $A_{k|4}$ is a projector satisfying $\sum\limits_{k}A_{k|4}=\mathbb{I}$. Let Alice's measurement on her system produces the outcomes with same probability, and having no pattern. The question is whether such an unpredictability of the outcomes are genuine or someone (say, an adversary Eve) may be able to guess Alice's outcomes. To examine the presence of Eve, the unpredictability of Alice's outcomes has to be certified in a device-independent way. Alice then performs certain tests for the device-independent certification of Eve's guessing probabilities of her outcomes. This in turn requires the certification of three-outcome POVMs and in that case the a upper bound of Eve's guessing probability needs to be $1/3$. First test is the maximization of the Bell expression $\mathbb{B}_{3}$ which has already been shown and second test is the minimization of the probability of other events. 

We first show that a modified Bell expression of $\mathcal{B}_{3}$ can  be used to self-test three outcome POVMs without invoking the role of Eve. However, while demonstrating the certification of randomness we explicitly consider Eve's role. Following Acin \emph{et al.}\cite{acin16},  let us define a modified Bell expression $\mathcal{B}^{\prime}_{3}$  as
\begin{align}
\label{povm}
	\mathcal{B}^{\prime}_{3}=\mathcal{B}_{3}-\alpha\sum_{k=1}^{3} P(k,+|x=4,y=k)
\end{align}
where $\alpha$ is strictly positive. As the last term in the right hand side is always negative then  both $\mathcal{B}_{3}^{\prime}$ and $\mathcal{B}_{3}$ has same classical upper bound. Equivalently, $(\mathcal{B}^{\prime}_{3})_{Q}$ cannot be larger than $(\mathcal{B}_{3})_{Q}^{opt}=6$ and there is only way to obtain equality when  the probability  $P(k,+|x=4,y=k))=Tr[\left({A_{k|4}}\otimes \Pi_{B_{k}}^{+} \right)\rho_{AB}]$ equals zero, for every $k$. Here $\Pi_{B_{k}}^{+}$ is the projector corresponding to the Bob's observables. This is possible when the POVM elements $\{A_{k|4}\}$ is anti-aligned with three $\Pi_{B_{k}}^{+}$ of Bob. Since $\Pi_{B_{k}}^{+}$ are certified by the optimal value of $(\mathcal{B}_{3})_{Q}^{opt}=6$ and they are positive projectors of the trine-set of observables, we can then have $A_{k|4}=\frac{2}{3} \Pi_{A_{k}}^{-}=\frac{1}{3}(\mathbb{I}-A_{k})$ with $k=1,2,3$. Thus the modified Bell expression $(\mathcal{B}^{\prime}_{3})_{Q}^{opt}=6$ certifies the three-outcome POVMs. If Alice performs the POVM $A_{4}=\{A_{k|4}\}$ on  her subsystem then the probability of each outcome provides the same probability $1/3$ and generate the $log_{2}3$ bit of local randomness.

However, in general in randomness certification protocols Eve's role is crucial. Her strategy may be to use a POVM $F=\{F_{k}\}$ so that she can model her measurement in  a way that whenever Alice obtain the outcome $k$ she perfectly guesses that outcome. Then Eve's probability of perfectly guessing Alice's outcome is given by \cite{acin16}

\begin{align}
	G=\underset{{F}}{max} \sum\limits_{k} P(k,k|A_{k|3}, F)
\end{align}

If the Eve's guessing probability is found to be $G = 1/3$ then the unpredictability of Alice's outcomes is certified. In order to certify this in device-indepndent way, in general one can write a family of qubit POVMs operators $A_{4}=\{A_{k|4}\}$ as
\begin{align}
	A_{k|4}=\gamma_{k}^{0}\mathbb{I}+\gamma_{k}^{1} \sigma_z+\gamma_{k}^{2} \sigma_y+\gamma_{k}^{4} \sigma_x
\end{align}
where $\sigma_x,\sigma_z,\sigma_y$ are the Pauli operators. It can be readily checked that to satisfy $A_{k|4}=\frac{2}{3} \Pi_{A_{k}}^{-}$  the coefficients $\gamma_{k}^{j}$ with $j=0,1,2,3$ takes the form

\begin{eqnarray}
	\gamma_{k}^{0}&=&P(k|A_{k|4}); \	\gamma_{k}^{1}= E_{k|3,1}\ ;		\gamma_{k}^{2}= \sum_{j=1}^{3} E_{b|3,j}\\
		\gamma_{k}^{3}&=&\frac{1}{\sqrt{3}}\left(E_{k|3,2}-E_{k|3,3}\right)
\end{eqnarray}
where $E_{k|x,y}=\sum_{k} k P(k,b|x,y)  $. 

Let us now consider the action of Eve here. In non-local guessing game, Eve tries to guess the Alice's output with highest possible accuracy, as summarized in Sec. II. Thus the inclusion of Eve's system modify the isometry as $\Phi: \mathcal {H}_{A}\otimes {H}_{B}\otimes {H}_{E} \rightarrow (\mathcal {H}_{A}\otimes \mathcal {H}_{A^{\prime}}) \otimes ({H}_{B}\otimes \mathcal {H}_{B^{\prime}}) \otimes {H}_{E}$ , so that, $\Phi (|\psi_{ABE}\rangle \otimes |00\rangle_{A^{\prime}B^{\prime}})=|\chi\rangle_{ABE} \otimes |\psi^{+}\rangle _{A^{\prime} B^{\prime}}$. Now, on the support of $\mathcal{H}_{A}\otimes \mathcal{H}_{{A}^{\prime}}$, the each element $\{A_{k|4}\}$ of $A_4$ can be represented by an operator $\tilde{A}_{k|4} \in \mathcal{H}_{A}\otimes \mathcal{H}_{{A}^{\prime}}$, so that 
\begin{align}
	\tilde{A}_{k|4}=\sum_{j=0}^{3} \tilde{A}_{k|4}^{j}\otimes \sigma_{j}
	\end{align}
where $\sigma_{0}=\mathbb{I}$ and $j=1,2,3$ corresponds to Pauli operators. From the self-testing relations given by Eq. (\ref{sf12}), we have 
\begin{eqnarray}
\nonumber
	\gamma_{k}^{0}&=& \langle \psi|  \tilde{A}_{k|4}\otimes \mathbb{I}|\psi\rangle_{}=\langle\chi | \tilde{A}_{k|4}^{0} |\chi\rangle\\
	\nonumber
	\gamma_{k}^{1}&=&\langle \psi| \tilde{A}_{k|4}\otimes B_1|\psi\rangle_{}=\langle\chi | \tilde{A}_{k|4}^{1} |\chi\rangle\\
		\gamma_{k}^{2}&=& \langle \psi| \tilde{A}_{k|4}\otimes(B_1+B_2+B_3)|\psi\rangle_{}=\langle\chi | \tilde{A}_{k|4}^{2} |\chi\rangle\\
		\nonumber
	\gamma_{k}^{3}&=&\frac{1}{3}\langle \psi| \tilde{A}_{k|4}\otimes\otimes B_{2}-B_{3}|\psi\rangle_{}=\langle\chi | \tilde{A}_{k|4}^{3} |\chi\rangle
\end{eqnarray}
Here $|\psi\rangle\equiv |\chi\rangle_{ABE}\otimes |\psi^{+}\rangle_{A^{\prime} B^{\prime}}$. This thus self test the three-outcome extremal qubit POVMs $A_{k|4}=\frac{2}{3} \Pi_{A_{k}}^{-}$ with $k=1,2,3$. 

We now proceed to generate the certified randomness of more than one bit, builded upon the work by Acin \emph{et al.}\cite{acin16}. This is to show that for a quantum state $|\psi\rangle_{ABE}$ shared by Alice, Bob and Eve, and for $A_x$ and $B_y$ ($x,y=1,2,3$) local Alice and Bob respectively and a POVMs $\{F_{k}\}$ is local to Eve. If the optimal violation of $\mathcal{B}_{3}$ is obtained, this then certifies the local guessing probability $G=1/3$.  In order to showing this, by assuming a normalized state $|\phi^{k}_{{A^{\prime}}}\rangle=F_{k}|\chi\rangle/ \sqrt{q_{k}}$ and without loss of generality by taking Eve's measurement $F_k$ is projective, we can write

\begin{eqnarray}
\label{povm1}
	\gamma_{k}^{j}&=& \sum\limits_{k^{\prime}}\langle\chi |F_{k^{\prime}} \tilde{A}_{k|4}^{j} F_{k^{\prime}}|\chi\rangle\\
		&=& \sum\limits_{{k^{\prime}}} q_{{k^{\prime}}}\langle \phi^{{k^{\prime}}}_{{A^{\prime}}} | \tilde{A}_{k|4}^{j}|\phi^{{k^{\prime}}}_{{A^{\prime}}}\rangle =\sum\limits_{{k^{\prime}}} q_{{k^{\prime}}} \beta_{{k}}^{j;{k^{\prime}}}
		\end{eqnarray}
		where $k= 1,2,3$. This can be interpreted as a convex combination of original POVMs $\{A_{k|4}\}$ in terms of the POVMs $\{\tilde{A}_{k|4}\}$ with respective weight $q_{k}$, so that $A_{k|4}=\sum\limits_{k^{\prime}}\sum\limits_{j} q_{k^{\prime}} \beta_{k}^{j,k^{\prime}} \sigma_{j}$.  But since $A_{k|4}$ is extremal, we have $\beta_{k}^{j,k^{\prime}}=\gamma_{k}^{j}$ for all $k^{\prime}$. This then imply that $\beta_{k}^{0,k}=\gamma_{k}^{0}=1/3$ for all $k$. We then have the local guessing probability

\begin{eqnarray}
G&=&\sum\limits_{k} P(k,k|A_{k|4}, F_k)= \sum\limits_{k} \langle \psi | \tilde{A}_{k|4}F_{k}|\psi\rangle \\
\nonumber
&=& \sum\limits_{k}\langle \chi |\tilde{A}_{k|4}^{0} F_{k}|\chi\rangle= \sum\limits_{k} q_{k} \beta_{k}^{0,k}=1/3			
		\end{eqnarray}
		
 This then certify $H_{\infty}=log_{2} 3$ bit of randomness from one entanglement bit.  It is proven by D'Ariano \emph{et al.} \cite{dariano} that there exists  $d^2$ number of extremal POVMs for $d$ dimensional space. Using this fact, Acin \emph{et al.}\cite{acin16} argued that at most $2 \ log_{2}\ d$ (and $4 \ log_{2} \ d$ bits of local (global) randomness can be certified from an entangled state of dimension $\mathcal{C}^{d}\otimes \mathcal{C}^{d}$.  For the case of qubit system, Acin \emph{et al.}\cite{acin16} demonstrated an interesting protocol to certify two bit of local randomness based on a simultaneous maximal quantum violation of three Clauser-Horne-Shimony-Holt inequalities. They \cite{acin16} had conjectured that the maximum quantum violation of Gisin's  elegant Bell inequality''  \cite{gisin} can also be used to certify two bit of randomness in a device-independent way which is proved by Andersson et al. \cite{and18} by providing the self-testing properties of elegant Bell inequality\cite{and17}. Here we use less number of observables for Alice and Bob and device-independently certify $log_{2} 3$ bit of local randomness.

\section{Generalization of the  game for any arbitrary odd $n$}

We now generalize the parity-oblivious communication game where Alice and Bob hold input $x,y \in \{1,2,...n\}$ and outputs are $a,b\in \{0,1\}$. Alice prepares $2n$ input states $x^{i}\in (x,a)\equiv \{1,2...n\}\times \{0,1\}$ and sends  to Bob. We consider uniform distribution of inputs of Alice and also for Bob, so that,  $p_{A}(x)=p_{B}(y)=1/n$.  The condition of winning the game remains same as before, i.e., $b=\delta_{x,y}\oplus_{2} a$.  In such a case the success probability is given by

\begin{eqnarray}
	\mathbb{P}_{n}= \frac{1}{2}+\frac{\langle \mathcal{B}_{n}\rangle}{2n^2}
\end{eqnarray}
where $\mathcal{B}_{n}$ is the Bell expression is given by

\begin{align}
\label{bni}
	\mathcal{B}_{n}=\sum_{x,y=1 x\neq y}^{n}  A_{n,x} \otimes B_{n,y} -\sum_{x,y=1 ; x= y}^{n} A_{n,x} \otimes B_{n,y}
\end{align}
In order to find the quantum upper bound of the Bell expression $\mathcal{B}_{n}$,  we again use the SOS approach. This is equivalent to showing that there is a positive semi-definite operator $\gamma_{n} \geq 0$, where

  \begin{align}
\label {gamman}
	 \gamma_{n} =\frac{1}{2} \sum_{y=1}^{3} \omega_{n,y} L_{n,y}^\dagger L_{n,y} 
 \end{align}
	where $ L_{n,y} $s are positive operators and polynomial functions of   $A_{n,x}$ and $B_{n,y}$. For the Bell expression given by Eq. (\ref{bni}), the operators $L_{n,y}$ can be written as

\begin{align}
\label{mni}
	L_{n,y}|\psi\rangle=\frac{1}{\omega_{n,y}}\sum\limits_{x=1}^{n} \alpha_{n}^{x,y} A_{n,x}|\psi\rangle -B_{n,y}|\psi\rangle
\end{align}
where $\alpha_{n}^{x,y}=1 (-1)$ when $x\neq y (x=y)$ and $	\omega_{n,y}=||\sum\limits_{x=1}^{n} \alpha_{n}^{x,y} A_{n,x}|\psi\rangle||$.  Plugging Eq. (\ref{mni}) into Eq. (\ref{gamman}) and by noting that $A_{n,x}^{\dagger} A_{n,x}=B_{n,y}^{\dagger} B_{n,y}=\mathbb{I} $, we get $\langle \gamma_{n}\rangle_{Q}=-(\mathcal{B}_{n})_{Q} + \sum\limits_{y=1}^{n}\omega_{n,y}$.  To obtain maximum value of $(\mathcal{B}_{n})_{Q}$ we can write 
\begin{align}
\label{max1}
max\left((\mathcal{B}_{n})_{Q}\right)\leq max\left(\sum\limits_{y=1}^{n}\omega_{n,y}\right) +max\left(-\langle \gamma_{n}\rangle_{Q}\right) 
\end{align}

For maximizing $\sum\limits_{y=1}^{n}\omega_{n,y}$ we again use the concavity inequality $	\sum\limits_{y=1}^{n}\omega_{n,y}\leq \sqrt{n \sum\limits_{y=1}^{n} \omega_{n,y}^{2}}$ where
\ba
	\nonumber
	\sum\limits_{y=1}^{n} \omega_{n,y}^{2}&=&n^{2}+\sum\limits_{y=1}^{n}\Bigg\langle\Bigg[ \left\{\alpha_{n}^{1,y} A_{n,1},\sum\limits_{x=2}^{n} \alpha_{n}^{x,y} A_{n,x}\right\} \\
	&+&\left\{\alpha_{n}^{2,y} A_{n,2},\sum\limits_{x\neq 2
	; x= 1}^{n} \alpha_{n}^{x,y} A_{n,x}\right\} \\
		\nonumber
	&+&....... \left\{\alpha_{n}^{n,y} A_{n,n},\sum\limits_{ x= 1}^{n-1} \alpha_{n}^{x,y} A_{n,x}\right\}\Bigg]\Bigg\rangle
\ea
which can be written in a simplified form as
\ba
	\sum\limits_{y=1}^{n} \omega_{n,y}^{2}=n^{2}+ (n-4)\langle \Delta_{n}\rangle 
	\ea
	where 
	
	\ba
\Delta_{n}&=&	\left\{A_{n,1}, \sum\limits_{x=2}^{n} A_{n,x}\right\} +\left\{A_{n,2}, \sum\limits_{x=3}^{n} A_{n,x}\right\} +.....\\
\nonumber
&&.....+\left\{A_{n,n-2}, \left( A_{n,n-1}+A_{n,n}\right)\right\}	 +\left\{A_{n,n-1},  A_{n,n}\right\}	
	\ea
	By considering $(\sum\limits_{x=1}^{n} A_{n,x})^{2}=n \mathbb{I} +\Delta_{n}$ for dichotomic observables that satisfies parity-oblivious condition, the optimal quantum value of $\langle \Delta_{n}\rangle=-n$. This in turn provides $max\left(\sum\limits_{y=1}^{n} \omega_{n,y}\right)= 2n$. 
	
	Since $\gamma_{n}$ is a positive operator, $\max\left(-\langle\gamma_{n}\rangle_{Q}\right)= 0$ for every $|\psi\rangle$, providing $\forall y$, $L_{n,y}|\psi\rangle=0$, i.e.,  
\begin{eqnarray}
\label{condan}
	\sum\limits_{x=1}^{n} \alpha_{n}^{x,y} A_{n,x}|\psi\rangle= \omega_{n,y} B_{n,y} |\psi\rangle
\end{eqnarray}
	Putting altogether, from Eq.(\ref{max1}), we have
\begin{align}
\label{sss}
(\mathcal{B})_{Q}^{opt} =2n
\end{align}
 One of the choices ofthe observables can be found for qubit system which are the following.

\ba
\label{choicen}
\nonumber 
	 &A_{n,1}=\sigma_{z}; \ \ 	\{A_{n, i}\}_{i=2,...\frac{n+1}{2}}=\nu_{n,i}\sigma_{x} -\beta_{n,i}\sigma_{y} - \frac{\sigma_{z}}{(n-1)}\\
		&\{A_{n, j}\}_{j=\frac{n+3}{2}, ... n}=-\nu_{n,j}\sigma_{x} +\beta_{n,j}\sigma_{y} - \frac{\sigma_{z}}{(n-1)}
\ea
 with $\nu_{n,i}^{2} +\beta_{n,i}^{2} +\frac{1}{(n-1)^2}=1$. Also, $\nu_{n,i}=\nu_{n,j}$ and $\beta_{n,i}=\beta_{n,j}$ when $i=(n+1)/2$ and $j=(n+3)/2$. In quantum theory, such choices of observables satisfy 
 \begin{align}
\label{poo}
	 A_{n,1} +\sum_{i=2}^{\frac{n-1}{2}} A_{n, i}+\sum_{j=\frac{n+1}{2}}^{n} A_{n, j}=0
 \end{align}
	and consequently the corresponding projectors satisfy the relation 
\ba
\label{nneq}
	\dfrac{2}{n} \left(P_{A,1}^{+(-)} +\sum_{i=2}^{\frac{n-1}{2}} P_{A_{n, i}}^{+(-)}+\sum_{j=\frac{n+1}{2}}^{n}  P_{A_{n, j}}^{+(-)}\right)= \mathbb{I}
\ea
This in turn satisfy the parity-oblivious condition of the game. The required state is a maximally entangled state given by Eq. (\ref{enstate}). Using Eq. (\ref{condan}) we can obtain Bob's choice of observables for optimal violation. It can be seen that Bob's observables satisfy the same condition of Alice as given by Eq. (\ref{poo}). 

Using the parity-oblivious condition given by Eq. (\ref{nneq}), the preparation non-contextuality bound of $\mathcal{B}_{n}$ can be derived which is $(\mathcal{B}_{n})_{pnc} \leq 2n-2$ thereby violated by quantum theory. The proof is similar to the case for $n=3$ which can be seen as follows. Putting the condition given in Eq. (\ref{poo}) to the generalized Bell expression in Eq. (\ref{bni}) we have $\mathcal{B}_{n}=2\sum\limits_{x=y=1}^{n} A_{x}B_{y}$. Using the Eq. (\ref{poo}) again it is straightforward to derive the aforementioned upper bound of  $(\mathcal{B}_{n})_{pnc}$ in a preparation non-contextual model. However, there is flexibility to choice a different form of the observables satisfying the same condition of Eq. (\ref{nneq}). In Appendix B we provide a self-testing protocols for certifying the maximally entangled state and observables based on the optimal quantum value of generalized Bell expression in Eq. (\ref{bni}).

Now, one may be wondering whether it is possible to certify $n$-outcome POVM and $log_{2}n $ bit of randomness for arbitrary (odd) $n$ scenario similar to the $n=3$. We discuss that this is not the case. Following $n=3$ case, let us consider the action of the $(n+1)^{th}$ measurement  $A_{n,n+1}$ of Alice  which is a $n$-outcome qubit POVMs  $\{A_{k|n+1}\}$. As already used earlier, a shifted Bell expression of Eq. (\ref{bni}) can be written as 
\begin{align}
\label{modbn}
	\mathcal{B}_{n}^{\prime}=\mathcal{B}_{n}-\alpha^{\prime}\sum_{k=1}^{n} P(k,+|x=n+1,y=k)
\end{align}
where $\alpha^{\prime}$ is strictly positive quantity. As the last term in the right hand side is always non-negative the preparation non-contextual bounds of both $\mathcal{B}_{n}^{\prime}$ and $\mathcal{B}_{n}$ remains $2n -2$. Similarly, the quantum value of $\mathcal{B}^{\prime}_{n}$ cannot exceed $2n$. There is only way to obtain maximum value of $(\mathcal{B}^{\prime}_{n})_{Q}$ when for every $k$ the probability  $P(a=k,b=+|x=n+1,y=k))$ equals zero. This is possible when the POVM elements of the measurement $\{A_{k|n+1}\}$ is anti-aligned with $n$ projective measurements of Bob's side. Such POVM elements can then be written as $A_{k|n+1}=\frac{2}{n}\Pi_{A_{k}}^{-}$ with $\frac{2}{n}\sum\limits_{k=1}^{n} \Pi_{A_{k}}^{-}=\mathbb{I}$. We already have the parity-oblivious condition in Eq. (\ref{nneq}) where such condition is satisfied for qubit observables. 

Note here that every element of POVM $\{A_{k|n+1}\}$ is effectively a projector. One may expect to certify of $n$ outcome qubit POVMs as in $n=3$ case, and consequently $log_{2}{n}$ bit of randomness. However, $\{A_{k|n+1}\}$ is not extremal set of POVMs if (odd) $n>3$.   It is known that the extremal POVMs for qubit systems have at most four outcomes and non-extremal POVMs can be simulated by convex combination of extremal POVMs \cite{oz18}. This indicates that the certification of unbounded randomness is not possible by using our generalized version of the game for arbitrary $n$. However, the optimal quantum success probability for arbitrary $n$ case enables device-independent self-testing of entangled state and measurements.

\section{Summary and conclusions}
In summary, we  provided an interesting oblivious communication game  played between two parties, Alice and Bob who receive arbitrary $n$ (odd) number of inputs. In particular, we provided an entanglement-assisted parity-oblivious game where Alice is allowed to communicate any amount of information but that should not reveal the parity information of the inputs to Bob. Such an oblivious condition in an operational theory implies obliviousness at the level of ontic states for preparation non-contextual ontological model \cite{spek09}. We showed that given any arbitrary $n$ the success probability of our game is solely dependent on a relevant Bell expression $\mathcal{B}_{n}$.  We demonstrated that the upper bound $\mathcal{B}_{n}$ can be reduced from the trivial case (the local bound), which we termed here as non-trivial preparation non-contextual bound. Aforementioned Bell expression is optimized by using SOS approach and it is found that a two-qubit maximally entangled state and an interesting set of observables in qubit system will suffice the purpose of optimization. Interestingly, the set of observables leading the optimal value satisfy the required parity-oblivious condition of the Alice's inputs. This provide a functional relationship between Alice choice of observables which in turn reduce the local bound to non-trivial preparation non-contextual bound of $\mathcal{B}_{n}$. Thus, for specific choice of states and measurements, it is possible that optimal quantum value of $(\mathcal{B}_{n})_{Q}^{opt}$ may not be enough to exhibit non-locality but the non-classicality in the form of non-trivial preparation contextuality may be demonstrated. 

Using the optimal quantum value $(\mathcal{B}_{3})_{Q}^{opt}=6$ for $n=3$ case, we first demonstrated the self-testing of two-qubit maximally entangled state and trine-spin observables. We used it to certify the three-outcome extremal POVMs which in turn enables us to certify the $log_{2} 3$ bit of local randomness.  Further, we generalized our scheme for any arbitrary (odd) number of inputs $n$ of Alice and Bob and demonstrated that optimal quantum value of Bell expression $(\mathcal{B}_{n})_{Q}^{opt}$ can be obtained for qubit system local to Alice and Bob. One may intend to exmine the possibility of certifying $n$-outcome POVMs by using a modified version of the Bell expression and its optimal quantum value $(\mathcal{B}_{n}^{\prime})_{Q}^{opt}$. Since extremal POVMs for qubit systems have at most four outcomes and non-extremal POVMs can be simulated by convex combination of extremal
measurements then the certification of $n$-outcome POVM is not possible. Finally, we may remark that since the generalized Bell expression can be optimized for qubit system, then the parity-oblivious communication game presented here can be tested by using the existing technologies. 
 \section*{Acknowledgments}
Author acknowledge the support from the project DST/ICPS/QuEST/Theme 1/2019/4.
\begin{widetext}
\appendix
\section*{Appendix}
\section{Self-testing of state and measurements based on the optimal value of $\mathcal{B}_{3}$}
We first provide the detailed derivation of self-testing of maximally entangled state and trine-set of spin observables for $n=3$ scenario based on the optimal quantum violation of  Bell's inequality Eq.(\ref{bi}). Specifically, we prove here the Eqs. (\ref{sf11}-\ref{sf14}) by using the self-testing circuit in Figure 1.  The self-testing relations given by Eqs. (\ref{st1}-\ref{st2}) are derived as follows. We define $Z_{A}= A_1$, $\tilde{X}_A = (A_3 -A_2)/\sqrt{3}$, $Z_B = -B_1 $ and $\tilde{X}_B = (B_2 -B_3)/\sqrt{3}$. From Eq. (18), the optimal quantum violation ensures that   $A_{1}B_{1}|\psi\rangle_{AB}=A_{2}B_{2}|\psi\rangle_{AB}=A_{3}B_{3}|\psi\rangle_{AB}=-|\psi\rangle_{AB}$.

Using $|\psi_{AB}\rangle = - A_{1}B_{1}|\psi_{AB}\rangle$ we have $A_{1}|\psi_{AB}\rangle = - B_{1}|\psi_{AB}\rangle$ and hence

\begin{align}
\label{a1}
	{Z}_A |\psi\rangle_{AB}={Z}_B|\psi\rangle_{AB}.
\end{align}

Similarly, we can write 
\begin{align}
\label{xx}
	\tilde{X}_{A} \tilde{X}_{B}|\psi\rangle_{AB}=\frac{1}{3}(A_3 B_2 - A_3 B_3 - A_2 B_2 +A_2 B_3)|\psi\rangle_{AB}= \frac{1}{3} (2+ A_3 B_2 + A_2 B_3)|\psi\rangle_{AB}
\end{align}
We first prove that $(A_3 B_2 + A_2 B_3)|\psi\rangle_{AB}=|\psi\rangle_{AB}$. Note that the optimal violation provides $A_1 +A_2 + A_{3}=0$ to hold for Alice's observables and $B_1 +B_2 + B_{3}=0$ for Bob's observables. Then by considering $A_{1}(B_{1}+B_{2}+B_{3})|\psi\rangle_{AB}=0$ we get $\left(A_{1} B_{2} + A_{1} B_{3}\right)|\psi\rangle_{AB}=|\psi\rangle_{AB}$. Similarly we get five more relations as $\left(A_{2} B_{1} + A_{2} B_{3}\right)|\psi\rangle_{AB}=|\psi\rangle_{AB}$, $\left(A_{3} B_{1} + A_{3} B_{2}\right)|\psi\rangle_{AB}=|\psi\rangle_{AB}$, $\left(A_{2} B_{1} + A_{3} B_{1}\right)|\psi\rangle_{AB}=|\psi\rangle_{AB}$, $\left(A_{1} B_{2} + A_{3} B_{2}\right)|\psi\rangle_{AB}=|\psi\rangle_{AB}$ and $\left(A_{2} B_{3} + A_{1} B_{3}\right)|\psi\rangle_{AB}=|\psi\rangle_{AB}$. Using those relations it can be proved that $\left(A_{3} B_{2} + A_{2} B_{3}\right)|\psi\rangle_{AB}= |\psi\rangle_{AB}$. In fact, it can be proved that each of $A_{x} B_{y}|\psi\rangle_{AB}$ is equal to each other for any $x$ \emph{not} equal to $y$.
 
From Eq. (\ref{xx}) we then have $\tilde{X}_{A} \tilde{X}_{B}|\psi\rangle_{AB} = |\psi\rangle_{AB}$, i.e., 
\begin{align}
	\label{a2}
	\tilde{X}_A |\psi\rangle_{AB}=\tilde{X}_B|\psi\rangle_{AB}.
\end{align}
Using the aforementioned relations it can be shown that 
\begin{align}
	\left({Z}_{A} \tilde{X}_{A} +\tilde{X}_{A} {Z}_{A} \right)|\psi\rangle_{AB}= (1/3)\left( A_{1} B_{3} +A_{3} B_{1} -A_{1} B_{2} -A_{2} B_{1}\right)|\psi\rangle_{AB}=0
\end{align}
 and 
\begin{align}
	\left({Z}_{B} \tilde{X}_{B} +\tilde{X}_{B} {Z}_{B} \right)|\psi\rangle_{AB}= (1/3)\left(B_{1}B_{3}+B_{3}B_{1} -B_{1} B_{2} -B_{2} B_{1}\right)|\psi\rangle_{AB}=0
\end{align}
 We then have 
\begin{align}
\label{a3}
\{{Z}_A, \tilde{X}_A\} |\psi\rangle_{AB} =\{{Z}_B, \tilde{X}_B\} |\psi\rangle_{AB}=0
	\end{align}
	Eqs. (\ref{a1}) , (\ref{a2}) and (\ref{a3}) are the the self-testing properties provided in the main text in Eqs. (\ref{st1}-\ref{st2}).
	
Using the isometry described in Figure 1 we can write
\begin{eqnarray}
\Phi(|\psi\rangle_{AB}\otimes |00\rangle_{A^{\prime}B^{\prime}})&=& \frac{1}{4}\Big[(1+Z_A)(1+Z_B)|\psi\rangle_{AB}|00\rangle+\tilde{X_{B}}(1+Z_A)(1-Z_B)|\psi\rangle_{AB}|01\rangle\\
\nonumber
&+&X_{A}(1-Z_A)(1+Z_B)|\psi\rangle_{AB}|10\rangle +\tilde{X_{A}}\tilde{X_{B}}(1-Z_A)(1-Z_B)|\psi\rangle_{AB}|11\rangle\Big]
\end{eqnarray}
which can be recast by using the self-testing properties given by Eqs.(\ref{st1}) and (\ref{st2}), as 
\begin{eqnarray}
\Phi(|\psi\rangle_{AB}\otimes |00\rangle_{A^{\prime}B^{\prime}})&=& \frac{1}{4}\Big[2(1+ Z_A)|\psi\rangle_{AB}\otimes \left(|00\rangle+|11\rangle\right)\Big]\\
\nonumber
&\equiv&\frac{(1+ Z_A)}{\sqrt{2}}|\psi\rangle_{AB}\otimes |\phi^{+}\rangle_{A^{\prime}B^{\prime}}
\end{eqnarray} 
Identifying $|\chi\rangle_{AB}  =\frac{(1+ Z_A)}{\sqrt{2}}|\psi\rangle_{AB}$ we have the Eq. (\ref{sf11}) in the main text. This implies the self-testing of two-qubit maximally entangled state using the optimal quantum value of the Bell expression in Eq. (\ref{bi}). 

Now for self-testing of measurements, we note that $B_1= -Z_{B}$, $B_2 = (\sqrt{3}/2)\tilde{X}_{B} +(1/2 ) Z_{B}$ and $B_{3}= -(\sqrt{3}/2)\tilde{X}_{B} +(1/2 ) Z_{B}$. Similar relations can be written for $A_{1}, A_{2}$ and $A_{3}$. It is then enough to demonstrate how the local isometry works for $Z_{B}$, $X_{B}$, $X_{A}$ and $Z_{A}$. We thus first show the following by using the self-testing circuit.
  \begin{eqnarray}
\label{asf11}
	\Phi(\tilde{X_{A}}|\psi\rangle_{AB}\otimes |00\rangle_{A^{\prime}B^{\prime}})&=& \frac{1}{4}\Big[(1+Z_A)\tilde{X_{A}}(1+Z_B)|\psi\rangle_{AB}|00\rangle +\tilde{X_{B}}(1+Z_A)\tilde{X_{A}}(1-Z_B)|\psi\rangle_{AB}|01\rangle\\
\nonumber
&+&\tilde{X_{A}}(1-Z_A)\tilde{X_{A}}(1+Z_B)|\psi\rangle_{AB}|10\rangle +\tilde{X_{A}}\tilde{X_{B}}(1-Z_A)\tilde{X_{A}}(1-Z_B)|\psi\rangle_{AB}|11\rangle\Big]
\end{eqnarray}
The first term $(1+Z_A)\tilde{X}_{A}(1+Z_B)|\psi\rangle_{AB}|00\rangle = \left(\tilde{X}_{A}+ \tilde{X}_{A}Z_{B} +Z_{A}\tilde{X}_{A}+Z_{A}\tilde{X}_{A}Z_{B}\right)|\psi\rangle|00\rangle$. Using $Z_{A}|\psi\rangle=Z_{B}|\psi\rangle$ and $\{Z_{A},\tilde{X}_{A}\}|\psi\rangle_{AB}=0$ we find $\left(\tilde{X}_{A}+ \{Z_{A},\tilde{X}_{A}\} -\tilde{X}_{A}\right) |\psi\rangle_{AB}|00\rangle=0$. The second term is given by $\left(\tilde{X}_{B}\tilde{X}_{A}-\tilde{X}_{B}\tilde{X}_{A} Z_{B}+\tilde{X}_{B} Z_{A}\tilde{X}_{A}-\tilde{X}_{B} Z_{A}\tilde{X}_{A} Z_{B}\right)|\psi\rangle_{AB}|01\rangle=\left(2 + Z_{A} +Z_{B}\right)|\psi\rangle_{AB}|01\rangle=2\left(1 + Z_{A}\right)|\psi\rangle_{AB}|01\rangle$. The third term $\left(\tilde{X}_{A}\tilde{X}_{A}+\tilde{X}_{A}Z_{A}\tilde{X}_{A} -\tilde{X}_{A} \tilde{X}_{A} Z_{B}-\tilde{X}_{A} Z_{A} \tilde{X}_{A}Z_{B} \right)|\psi\rangle_{AB}|10\rangle=2\left(1 + Z_{A}\right)|\psi\rangle_{AB}|01\rangle$. Similarly the fourth term  $\left(\tilde{X}_{A}\tilde{X}_{B}\tilde{X}_{A}-\tilde{X}_{A}\tilde{X}_{B} \tilde{X}_{A} Z_{B}-\tilde{X}_{A}\tilde{X}_{B}Z_{A}\tilde{X}_{A}+\tilde{X}_{A}\tilde{X}_{B}Z_{A}\tilde{X}_{A} Z_{B}\right)|\psi\rangle_{AB}|11\rangle$ can be written as $\left(\tilde{X}_{A}+\tilde{X}_{B} Z_{B}-\tilde{X}_{A}Z_{A}-\tilde{X}_{A}\tilde{X}_{B}\tilde{X}_{A} \right)|\psi\rangle_{AB}|11\rangle$ by using Eqs. (\ref{a1}) and (\ref{a2}). Further using them along with $\{Z_{A},\tilde{X}_{A}\}|\psi\rangle_{AB}=0$ from Eq(\ref{a3}), we have  $\left(\tilde{X}_{A}-\{\tilde{X}_{B},Z_{B}\} -\tilde{X}_{A}\right)|\psi\rangle_{AB}|11\rangle=0$. 

Using those, the Eq. (\ref{asf11}) can then be written as
\begin{eqnarray}
\label{asf111}
	\Phi(\tilde{X_{A}}|\psi\rangle_{AB}\otimes |00\rangle_{A^{\prime}B^{\prime}})= \frac{1+ Z_{A}}{\sqrt{2}}|\psi\rangle_{AB}\frac{\left(|01\rangle +|10\rangle\right)}{\sqrt{2}}=|\chi\rangle_{AB} \otimes \left(\sigma_{x}\otimes \mathbb{I}\right)|\phi^{+}\rangle_{A^{\prime}B^{\prime}}
\end{eqnarray}
Following the similar steps as above, we have
\begin{eqnarray}
	\label{asf10}
	\Phi(\tilde{X_{B}}|\psi\rangle_{AB}\otimes |00\rangle_{A^{\prime}B^{\prime}})&=& \frac{1}{4}\Big[(1+Z_A)(1+Z_B)\tilde{X_{B}}|\psi\rangle|00\rangle +\tilde{X_{B}}(1+Z_A)(1-Z_B)\tilde{X_{B}}|\psi\rangle|01\rangle\\
\nonumber
&+&X_{A}(1-Z_A)(1+Z_B)\tilde{X_{B}}|\psi\rangle|10\rangle +\tilde{X_{A}}\tilde{X_{B}}(1-Z_A)(1-Z_B)\tilde{X_{B}}|\psi\rangle|11\rangle\Big]\\
\nonumber
&\equiv& |\chi\rangle_{AB} \otimes \left(\mathbb{I}\otimes \sigma_{x}\right)|\phi^{+}\rangle_{A^{\prime}B^{\prime}}
\end{eqnarray}
and 
\begin{eqnarray}
\label{asf12}
	\Phi(Z_{A}|\psi\rangle_{AB}\otimes |00\rangle_{A^{\prime}B^{\prime}})&=& |\chi\rangle_{AB} \otimes \left(\sigma_{z}\otimes \mathbb{I}\right)|\phi^{+}\rangle_{A^{\prime}B^{\prime}}\\
	\label{asf13}
	\Phi(Z_{B}|\psi\rangle_{AB}\otimes |00\rangle_{A^{\prime}B^{\prime}})&=& |\chi\rangle_{AB} \otimes \left(\mathbb{I}\otimes\sigma_{z} \right)|\phi^{+}\rangle_{A^{\prime}B^{\prime}}
\end{eqnarray}

Using the results in Eqs. (\ref{asf10} -\ref{asf13}), it is straightforward to show that 
\begin{eqnarray}
\label{asf14}
	\Phi(B_{y}|\psi\rangle)&=&|\chi\rangle_{AB} \otimes (\mathbb{I}\otimes B_{y}^{\prime})|\phi^{+}\rangle_{A^{\prime}B^{\prime}}\\
	\label{asf15}
		\Phi(A_{x}|\psi\rangle)&=&|\chi\rangle_{AB} \otimes ( A_{x}^{\prime}\otimes\mathbb{I})|\phi^{+}\rangle_{A^{\prime}B^{\prime}}
		\end{eqnarray}
		Similarly, a few more steps are required to show
		\begin{eqnarray}
		\label{asf16}
	\Phi(A_{x}\otimes B_{y}|\psi\rangle)=|\chi\rangle_{AB} \otimes (A_{x}^{\prime}\otimes B_{y}^{\prime})|\phi^{+}\rangle_{A^{\prime}B^{\prime}}
			\end{eqnarray}
			We have thus self-tested the measurements and the Eqs. (\ref{asf14}-\ref{asf16}) are the Eqs. (\ref{sf12}-\ref{sf14}) in the main text.
\section{Self-testing for the case of $n=5$}
 We first note that there are different forms of  choices of observable available that satisfies the oblivious condition given by Eq. (\ref{poo}). One of the choices are presented in Eq. (\ref{choicen}). We can choose another set of observable by keeping the parity oblivious condition and consequently the functional relation between the observables of Alice and Bob intact. As already mentioned in the main text, such a functional relation fixes the optimal quantum value of the Bell expression Eq. (\ref{bni}). We provide the self-testing scheme for $n=5$ which can be straightforwardly generalized for any odd $n$. The alternative choices for $n=5$ are the following
\ba
	A_{5,1}&=&\sigma_{z}; \ \ 
	 	A_{5, 2}=\nu_{5,1}\sigma_{x} -\beta_{5,1}\sigma_{y} - \frac{\sigma_{z}}{4}; \ \ 		A_{5, 3}=-\nu_{5,1}\sigma_{x} -\beta_{5,1}\sigma_{y} - \frac{\sigma_{z}}{4}\\
		\nonumber 
		A_{5, 4}&=&\nu_{5,1}\sigma_{x} +\beta_{5,1}\sigma_{y} - \frac{\sigma_{z}}{4}; \ \ 
		A_{5, 5}=-\nu_{5,1}\sigma_{x} +\beta_{5,1}\sigma_{y} - \frac{\sigma_{z}}{4}	
\ea
satisfying the parity-oblivious condition $\sum\limits_{n=1}^{5} A_{5,n}=0$. Here $|\nu_{5,n}|^{2} +|\beta_{5,n}|^{2}+\frac{1}{16}=1$.

\begin{figure}[h]
  \includegraphics[width=10 cm,height=7 cm]{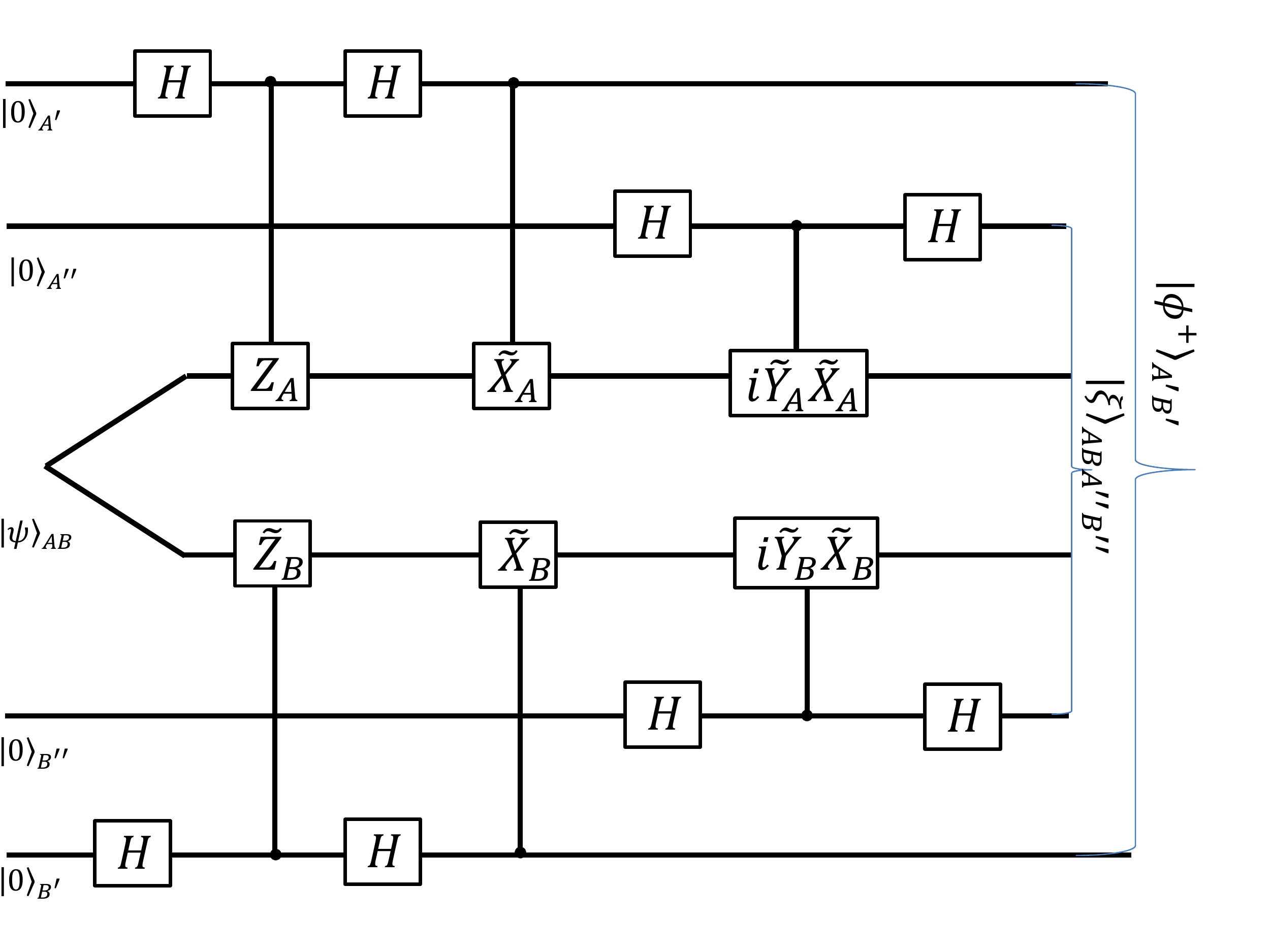}
 \centering
  \caption{(color online) Self-testing scheme based on the optimal quantum value of $\mathbb{B}_{5}$ which can be obtained by putting $n=5$ in Eq. (\ref{bi}). Details of SWAP-isometry are given in the  text.}
  \label{fig2}
\end{figure}
Following the usual SWAP circuit in Figure \ref{fig2} , we show that the maximum violation of $\mathcal{B}_{5}$ implies the existence of an isometry $\Phi$ so that $\Phi: \mathcal{H}_{A}\otimes \mathcal{H}_{B} \rightarrow (\mathcal{H}_{A}\otimes \mathcal{H}_{A^{\prime}}\otimes \mathcal{H}_{A^{\prime\prime}})\otimes(\mathcal{H}_{B}\otimes\mathcal{H}_{B^{\prime}}\otimes \mathcal{H}_{B^{\prime\prime}})$ and $|\psi\rangle_{AB}\rightarrow|\xi\rangle_{AB A^{\prime}B^{\prime}}\otimes |\phi^{+}\rangle_{A^{\prime\prime}B^{\prime\prime}}$.

Let us define the observables $Z_{A}= A_{5,1}$, $X_A = A_{5,2}-A_{5,3}+A_{5,4}-A_{5,5}$, $Y_A = -A_{5,2}-A_{5,3}+A_{5,4}+A_{5,5}$ along with  $Z_B = -B_{5,1} $, $X_B = -B_{5,2}+B_{5,3}-B_{5,4}+B_{5,5}$, $Y_B = -B_{5,2}-B_{5,3}+B_{5,4}+B_{5,5}$ satisfying following properties. The observables $X_A$, $X_B$, $Y_A$ and $Y_B$ may not be Hermitian but one can define the following Hermitian operators, for example, $\tilde{X}_A=X_A/ ||X_A||$. The self-testing relations  are the following.

\ba
\label{5st11}
	&& Z_A |\psi\rangle=Z_B|\psi\rangle;  \ \  \tilde{X}_A |\psi\rangle=\tilde{X}_B|\psi\rangle;\ \ \tilde{Y}_A |\psi\rangle=-\tilde{Y}_B|\psi\rangle\\
\label{5st21}
	&&\{Z_A, \tilde{X}_A\} |\psi\rangle \{\tilde{Y}_A, \tilde{X}_A\} |\psi\rangle =\{Z_A, \tilde{Y}_A\} |\psi\rangle =0\\
	\label{5st31}
		&&\{Z_B, \tilde{X}_B\} |\psi\rangle \{\tilde{Y}_B, \tilde{X}_B\} |\psi\rangle =\{Z_B, \tilde{Y}_B\} |\psi\rangle =0
	\ea
Using the above relations we can also have 
\ba
\label{yaxa}
\tilde{Y}_A \tilde{X}_A|\psi\rangle=\tilde{Y}_B \tilde{X}_B|\psi\rangle
\ea
which provides $\tilde{Y}_A \tilde{X}_A\tilde{Y}_B \tilde{X}_B|\psi\rangle =-|\psi\rangle$. Using the circuit in Figure \ref{fig2}, we first demonstrate the self-testing of the two-qubit maximally entangled state using the isometry $\Phi(|\psi\rangle_{AB}\otimes |00\rangle_{A^{\prime}B^{\prime}}\otimes |00\rangle_{A^{\prime\prime}B^{\prime\prime}})$.  

It is already shown in Appendix A that due to the action of the first two pairs of Hadamard operations and controlled gate operations of ${Z}_A $, $\tilde{X}_A $, ${Z}_B $ and $\tilde{X}_B $, the state evolves to 
\begin{eqnarray}
\label{b6}
 |\chi\rangle_{AB}\otimes |\phi^{+}\rangle_{A^{\prime}B^{\prime}}\otimes|00\rangle_{{A^{\prime\prime}B^{\prime\prime}}}
\end{eqnarray} 
where $|\chi\rangle_{AB}=\frac{(1+ Z_A)}{\sqrt{2}}|\psi\rangle_{AB}$. After the third pair of Hadamards the state is evolved to $
 |\xi\rangle_{AB}\otimes |\phi^{+}\rangle_{A^{\prime}B^{\prime}}\otimes|++\rangle_{{A^{\prime\prime}B^{\prime\prime}}}$ and applying the final set of controlled gates and finally using Eq. (\ref{yaxa}), the state in Eq.(\ref{b6}) evolves to  
\begin{eqnarray}
\label{ddd}
 |\chi\rangle_{AB}\otimes |\phi^{+}\rangle_{A^{\prime}B^{\prime}}\otimes \frac{1}{2}\left[|00\rangle_{{A^{\prime\prime}B^{\prime\prime}}}+i\tilde{Y}_{B}\tilde{X}_{B} |01\rangle_{{A^{\prime\prime}B^{\prime\prime}}}+i\tilde{Y}_{B}\tilde{X}_{B}|10\rangle_{{A^{\prime\prime}B^{\prime\prime}}}-|11\rangle_{{A^{\prime\prime}B^{\prime\prime}}}\right]
\end{eqnarray} 
Considering the action of the final pair of Hadamards, we find the self-testing of the entangled state is given by 
\begin{eqnarray}
\Phi(|\psi\rangle_{AB}\otimes |00\rangle_{A^{\prime}B^{\prime}}\otimes |00\rangle_{A^{\prime\prime}B^{\prime\prime}})= |\xi\rangle_{ABA^{\prime\prime}B^{\prime\prime}}\otimes |\phi^{+}\rangle_{A^{\prime}B^{\prime}}
\end{eqnarray} 
where $|\xi\rangle_{ABA^{\prime\prime}B^{\prime\prime}}$ is the so-called junk state is given by
\begin{align}
	|\xi\rangle_{ABA^{\prime\prime}B^{\prime\prime}}= \frac{1}{2}\left[|\chi\rangle_{AB}\otimes(\mathbb{I}+i\tilde{Y}_{A}\tilde{X}_{A})|00\rangle_{{A^{\prime\prime}B^{\prime\prime}}}+ |\chi\rangle_{AB}\otimes(\mathbb{I}-i\tilde{Y}_{A}\tilde{X}_{A})|11\rangle_{{A^{\prime\prime}B^{\prime\prime}}}\right]
\end{align}
 
Now, for the local unitary evolutions provide in Figure \ref{fig2} and using self-testing relations given by Eqs. (\ref{5st11} -\ref{5st21}) it is straightforward to demonstrate that  
\begin{eqnarray}
\Phi(\tilde{X}_{A}|\psi\rangle_{AB}\otimes |00\rangle_{A^{\prime}B^{\prime}}\otimes |00\rangle_{A^{\prime\prime}B^{\prime\prime}})= |\xi\rangle_{ABA^{\prime\prime}B^{\prime\prime}}\otimes (\mathbb{I}\otimes\sigma_{x})|\phi^{+}\rangle_{A^{\prime}B^{\prime}}
\end{eqnarray} 
\begin{eqnarray}
\Phi({Z}_{A}|\psi\rangle_{AB}\otimes |00\rangle_{A^{\prime}B^{\prime}}\otimes |00\rangle_{A^{\prime\prime}B^{\prime\prime}})= |\xi\rangle_{ABA^{\prime\prime}B^{\prime\prime}}\otimes (\mathbb{I}\otimes\sigma_{z})|\phi^{+}\rangle_{A^{\prime}B^{\prime}}
\end{eqnarray} 
\begin{eqnarray}
\Phi(\tilde{X}_{B}|\psi\rangle_{AB}\otimes |00\rangle_{A^{\prime}B^{\prime}}\otimes |00\rangle_{A^{\prime\prime}B^{\prime\prime}})= |\xi\rangle_{ABA^{\prime\prime}B^{\prime\prime}}\otimes (\sigma_{x}\otimes \mathbb{I})|\phi^{+}\rangle_{A^{\prime}B^{\prime}}
\end{eqnarray} 
\begin{eqnarray}
\Phi({Z}_{B}|\psi\rangle_{AB}\otimes |00\rangle_{A^{\prime}B^{\prime}}\otimes |00\rangle_{A^{\prime\prime}B^{\prime\prime}})= |\xi\rangle_{ABA^{\prime\prime}B^{\prime\prime}}\otimes (\sigma_{z}\otimes \mathbb{I})|\phi^{+}\rangle_{A^{\prime}B^{\prime}}
\end{eqnarray} 
However, for $\Phi(\tilde{Y}_{A}|\psi\rangle_{AB}\otimes |00\rangle_{A^{\prime}B^{\prime}}\otimes |00\rangle_{A^{\prime\prime}B^{\prime\prime}})$, the derivation is a little involved and we provide the sketch of the derivation. The state just before the final pair of Hadamards can be written as
\ba
\nonumber
\frac{1}{2}\left[|\chi\rangle_{AB}\otimes|00\rangle_{{A^{\prime\prime}B^{\prime\prime}}}+ i\tilde{Y}_{B}\tilde{X}_{B}|\chi\rangle_{AB}\otimes |01\rangle_{{A^{\prime\prime}B^{\prime\prime}}} +i\tilde{Y}_{A}\tilde{X}_{A}|\chi\rangle_{AB}\otimes |10\rangle_{{A^{\prime\prime}B^{\prime\prime}}} -\tilde{Y}_{A}\tilde{X}_{A}\tilde{Y}_{B}\tilde{X}_{B}|\chi\rangle_{AB}\otimes |11\rangle_{{A^{\prime\prime}B^{\prime\prime}}} \right] |\phi^{+}\rangle_{A^{\prime}B^{\prime}}\\
\ea  

Using Eq. (\ref{yaxa}) and after the action of final pair of Hadamards we finally have 
\ba
\nonumber
\Phi(\tilde{Y}_{A}|\psi\rangle_{AB}\otimes |00\rangle_{A^{\prime}B^{\prime}}\otimes |00\rangle_{A^{\prime\prime}B^{\prime\prime}})&=& \frac{1}{2}\left[|\chi\rangle_{AB}\otimes(\mathbb{I}+i\tilde{Y}_{A}\tilde{X}_{A})|00\rangle_{{A^{\prime\prime}B^{\prime\prime}}}- |\chi\rangle_{AB}\otimes(\mathbb{I}-i\tilde{Y}_{A}\tilde{X}_{A})|11\rangle_{{A^{\prime\prime}B^{\prime\prime}}}\right] \\
\nonumber
&\otimes& (\sigma_{y}\otimes  \mathbb{I})|\phi^{+}\rangle_{A^{\prime}B^{\prime}}\\
&\equiv&\sigma_{z}^{A^{\prime\prime}}|\xi\rangle_{ABA^{\prime\prime}B^{\prime\prime}}\otimes (\sigma_{y}\otimes  \mathbb{I})|\phi^{+}\rangle_{A^{\prime}B^{\prime}}
\ea
Similarly, following the earlier steps it can be easily seen that 
\begin{eqnarray}
\Phi(\tilde{Y}_{B}|\psi\rangle_{AB}\otimes |00\rangle_{A^{\prime}B^{\prime}}\otimes |00\rangle_{A^{\prime\prime}B^{\prime\prime}})= \sigma_{z}^{B^{\prime\prime}}|\xi\rangle_{AB A^{\prime\prime}B^{\prime\prime}}\otimes (\mathbb{I}\otimes \sigma_{y} )|\phi^{+}\rangle_{A^{\prime}B^{\prime}}
\end{eqnarray} 
In order to take care the conjugation issue involved with $\sigma_{y}$ the observable $\sigma_{z}^{B^{\prime\prime}}$ has to be suitably used.
\section{Sketch regarding the self-testing for the case of (odd) $n> 5$}
Now, for the self-testing purpose in the case of $n>5$ it is convenient to use a different set of observables which also satisfy the required parity oblivious condition in Eq. (\ref{poo}). We choose the following set of observables. When $n=4l+5$ with $l=0,1,2..$
\ba
\label{c1}
	&&A_{n,1}=\sigma_{z}; 	\{A_{n, 2+4m}\}=\nu_{n,m}\sigma_{x} -\beta_{n,m}\sigma_{y} - \frac{\sigma_{z}}{(n-1)}; 	 	\{A_{n, 3+4m}\}=-\nu_{n,m}\sigma_{x} -\beta_{n,m}\sigma_{y} - \frac{\sigma_{z}}{(n-1)}\\
		\nonumber
	 	&&\{A_{n, 4+4m}\}=\nu_{n,m}\sigma_{x} +\beta_{n,m}\sigma_{y} - \frac{\sigma_{z}}{(n-1)}; \{A_{n, 5+4m}\}=-\nu_{n,m}\sigma_{x} +\beta_{n,m}\sigma_{y} - \frac{\sigma_{z}}{(n-1)}
\ea
where $m=0,1,2,...n-5$ and $|\nu_{n,m}|^{2} +|\beta_{n,m}|^{2} + 1/{(n-1)^2}=1$.\\ 

For $n=4l+7$ along with the above set of observables we additionally require 
\ba
\label{c2}
A_{n, (n-1)}=\alpha_{n}^{\prime}\sigma_{x} -\beta_{n}^{\prime}\sigma_{y} - \frac{\sigma_{z}}{(n-1)}; \ \ \ \ A_{n, n}=-\alpha_{n}^{\prime}\sigma_{x} +\beta_{n}^{\prime}\sigma_{y} - \frac{\sigma_{z}}{(n-1)}
\ea
where $|\alpha_{n}^{\prime}|^{2} +|\beta_{n}^{\prime}|^{2} +1/{(n-1)^2}=1$. By suitably summing and subtracting the observables in Eqs. (\ref{c1} - \ref{c2}) one obtains $Z_{A}$, $Z_{B}$, $X_{A}$, $X_{B}$, $Y_{A}$ and $Y_{B}$ which will provide the self-testing relations in Eqs. (\ref{5st11}-\ref{yaxa}).

\end{widetext}

\end{document}